\documentclass{vgtc}                          

\ifpdf
  \pdfoutput=1\relax                   
  \usepackage{graphicx}                
  \DeclareGraphicsExtensions{.pdf,.png,.jpg,.jpeg} 
\else
  \usepackage{graphicx}                
  \DeclareGraphicsExtensions{.eps}     
\fi%

\graphicspath{{figures/}{pictures/}{images/}{./}} 

\usepackage{microtype}                 
\PassOptionsToPackage{warn}{textcomp}  
\usepackage{textcomp}                  
\usepackage{mathptmx}                  
\usepackage{times}                     
\usepackage{cite}                      
\usepackage{tabu}                      
\usepackage{booktabs}                  

\usepackage[table,xcdraw]{xcolor}
\PassOptionsToPackage{table,xcdraw}{xcolor}
\PassOptionsToPackage{hyphens}{url}
\usepackage{multirow}
\usepackage{enumitem}
\usepackage{booktabs}
\usepackage{hyperref} 
\usepackage{fancyhdr,graphicx,amsmath,amssymb}
\usepackage{caption}
\usepackage{subcaption}
\usepackage{soul} 
\include{pythonlisting}
\newif\ifnotes
\notestrue

\onlineid{1098}

\vgtccategory{Application or Design Study}

\title{Streamlining Visualization Authoring in D3 Through User-Driven Templates}

\author{Hannah Bako\thanks{e-mail: hbako@cs.umd.edu}\\ %
        \scriptsize University of Maryland%
\and Alisha Varma\thanks{e-mail:alishav@terpmail.umd.edu}\\ %
     \scriptsize University of Maryland %
\and Anuoluwapo Faboro\thanks{e-mail:afaboro@terpmail.umd.edu}\\ %
     \scriptsize University of Maryland %
\and Mahreen Haider\thanks{email:mhaider1@terpmail.umd.edu}\\ %
     \scriptsize University of Maryland%
\and Favour Nerrise\thanks{email:fnerrise@terpmail.umd.edu}\\ %
     \scriptsize University of Maryland %
\and Bissaka Kenah\thanks{email:bkenah@terpmail.umd.edu}\\ %
     \scriptsize University of Maryland %
\and Leilani Battle\thanks{e-mail:leibatt@cs.washington.edu}\\ %
     \parbox{1.4in}{\scriptsize \centering University of Washington}}

\abstract{
\vspace{-1mm}
D3 is arguably the most popular tool for implementing web-based visualizations. Yet D3 has a steep learning curve that may hinder it's adoption and continued use. To simplify the process of programming D3 visualizations, we must first understand the space of implementation practices that D3 users engage in. We present a qualitative analysis of 2500 D3 visualizations and their corresponding implementations. We find that 5 visualization types (Bar Charts, Geomaps, Line Charts, Scatterplots and Force Directed Graphs) account for 80\% of D3 visualizations found in our corpus. While implementation styles vary slightly across designs, the underlying code structure for all visualization types remains the same; presenting an opportunity for code reuse. Using our corpus of D3 examples, we synthesize reusable code templates for eight popular D3 visualization types and share them in our open source repository. Based on our results, we discuss design considerations for leveraging users' implementation patterns to reduce visualization design effort through design templates and auto-generated code recommendations.

} 

\CCScatlist{
    \CCScatTwelve{Human-centered computing}{Visu\-al\-iza\-tion}{Visualization systems and tools}{Visualization toolkits};
}

\begin{document}

\firstsection{Introduction}

\maketitle

\vspace{-1mm}
Visualization languages are popular tool for creating interactive visualizations~\cite{mei2018design}, where D3~\cite{bostock2011d3} is one of the most well-known and expressive library for creating web based visualizations~\cite{battle2018beagle,hogue2020searchingd3,harper2014deconstructingd3, harper2018convertingd3}. However, D3 is notoriously complex, often requiring extensive programming experience to use~\cite{mei2018design,satyanarayan2019critical}.
Though simpler languages (e.g., Vega-Lite~\cite{satyanarayan2016vega}) and direct manipulation tools (e.g., Data Illustrator~\cite{liu2018dataIll} and Lyra~\cite{carr2014lyra,zong2020lyra}) are available, they do not solve the user's problem if their interests are specific to D3. For example, the user may be required to use D3 at work, or they might be interested in adapting a specific D3 example found online that is not supported by less expressive tools. These users can't pivot to simpler tools or languages, and still need easier ways to program D3 visualizations.

However, easing the D3 implementation process requires that we first gain an understanding of \emph{what visual and interactive design patterns people frequently use when programming D3 visualizations}. Such information can explain users' goals when creating a visualization and what they might struggle with in the process. 
A plethora of D3 examples are available on repositories such as Bl.ocks.org~\cite{blocks:2016} and GitHub~\cite{GitHub:2008} and services like Observable~\cite{observable:2016}, which could enrich our understanding of users' implementation processes. 

\begin{figure}[tbp]
\centering
\includegraphics[width=0.9\columnwidth, clip]{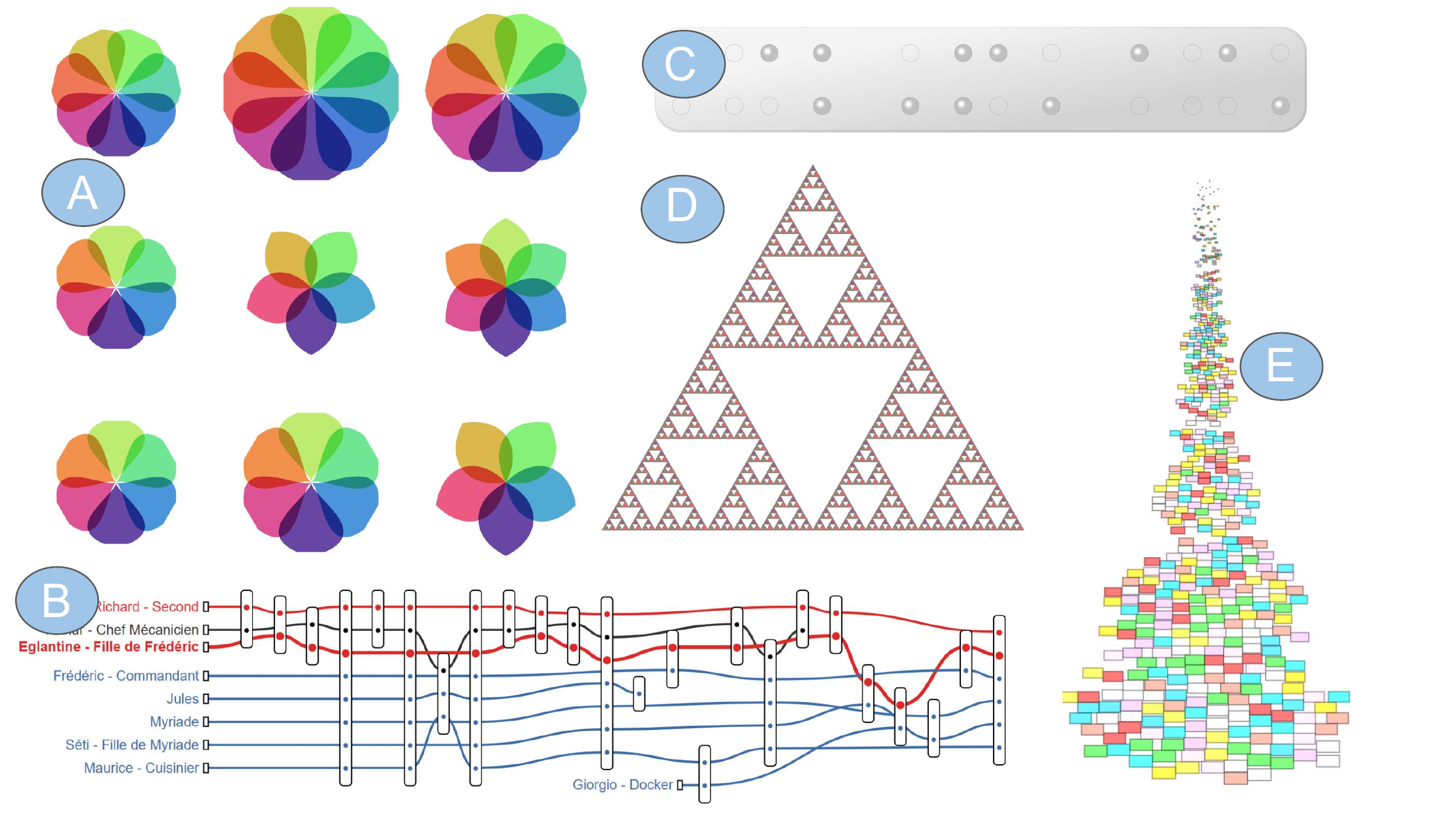}
\vspace{-5mm}
\caption{Examples of bespoke visualizations from our analysis. (A) renders the number of IMDB votes and corresponding ratings of movies in a movies dataset. (B) is a narrative chart of scenes from Star Wars: Episode IV. (C) visualizes a braille clock, (D) is a D3 rendering of Sierpinski Charlet, and (E) is a rendering of bounding box collisions using D3's force simulation.}
\label{fig:bespoke}
\vspace{-5mm}
\end{figure}

In this paper, we present an analysis of D3 code repositories and example galleries, highlighting common patterns to programming D3 visualizations.
We mined and qualitatively analyzed 2500 examples from GitHub, Bl.ocks.org, and Observable. To identify common strategies employed by D3 users, we analyze prevalent visualization types, the interactions (if any) used in these visualizations, and their corresponding code implementations. 

We find that standard visualizations (Bar charts, Geographic maps, Line Charts, Scatterplots and Graphs) account for over 80\% of the visualizations in our corpus, consistent with prior work~\cite{battle2018beagle}.
Furthermore, we observe a consistent pattern across our corpus: users' implementation strategies were largely the same within each visualization type.
We also find that interactions feature prominently within users' D3 implementations, and certain interaction types are commonly associated with specific visualization types. 
However a notable fraction of users also created \emph{bespoke} visualizations (see \autoref{fig:bespoke}),
suggesting quantifiable limits to common design patterns. 

To make our findings actionable, we synthesize visualization templates for eight popular D3 visualizations, which encapsulate the observed implementation practices of D3 users. 
These templates are available through our open source repository~\footnote{
\url{https://osf.io/k58bp/?view_only=72fa3798bbaa4263b5ad662b26a70cb3}.}. Informed by our observations, we discuss critical design considerations for future visualization programming tools that involve code reuse and user-driven recommendations for visualization and interaction design.

To summarize, we make the following contributions in this paper:
\begin{itemize}[nosep]
    \item We \textbf{analyze common visualization and implementation patterns} across 2500 D3 examples from three online repositories.
    \item We provide a suite of \textbf{general purpose templates} encapsulating common implementation strategies for the most popular D3 visualization and interaction types.
    \item We \textbf{discuss critical design considerations} for future visualization programming tools based on our findings.
\end{itemize}

\vspace{-1mm}
\section{Related Work} \label{sec:related-works}
\vspace{-1mm}
Our work builds on research in visualization authoring, interaction taxonomies, and templates. We review the relevant literature below.

\textbf{Visualization Authoring Tools:}
There are a plethora of tools for producing visualization designs, including language- (e.g~\cite{plotly2015,bostock2011d3,satyanarayan2016vega,satyanarayan2016reactivevega}) and GUI-based (e.g~\cite{stolte2002polaris, carr2014lyra, zong2020lyra,liu2018dataIll}) tools. Specification languages are expressive, but require considerable programming skills to use them well~\cite{vartak2017towards, dibia2019data2vis, hu2019vizml}. GUI-based tools eliminate the need for programming, but in return take control away from the user~\cite{horvitz1999principles} and still have a considerable learning curve~\cite{satyanarayan2019critical}.
Visualization users may be better supported by focusing on how to simplify the process of using visualization languages. To this end, we investigate users' implementation practices when using languages like D3.

\textbf{Visualization and Interaction Taxonomies:}
Interactions are critical because they allow users to elicit a deeper understanding of their data~\cite{dimara2020interactions}. 
Heer and Shneiderman provide a taxonomy of 12 interaction tasks grouped into three high-level categories: (1) Data \& View Specification tasks, (2) View Manipulation tasks and (3) Process \& Provenance tasks~\cite{heer2012interactivedynamics}. Brehmer and Munzner~\cite{brehmer2013typology} bridge the gap between high-level and low-level interaction task[s] classification by connecting complex high-level tasks as a sequence of simpler low-level tasks organized by the intent for the task being performed. To model the space of possible interactions, we incorporate tasks from Brehmer and Munzner's typology~\cite{brehmer2013typology} to analyze the space of interactions used in D3 visualizations.



\textbf{Templates:} Code reuse has been a strongly advocated and documented programming practice ~\cite{kim2004ethnographic, lin2017mining}.
Within the visualization community, some tools utilize templates to generate visualizations~\cite{McNutt2021Templates, d3.js/live, Mauri2017RawGraphs} and style templates~\cite{harper2018convertingd3}. 
Our work extends this line of research on using synthesized templates to support code reuse.
However, unlike past work that focuses on the underlying style structure of D3 charts~\cite{harper2018convertingd3}, our focus is on understanding what users consider to be \emph{intuitive implementation strategies} for D3 and extracting the commonalities across these strategies.
\begin{figure*}[t]
\begin{subfigure}{0.6\linewidth}
    \centering
    \includegraphics[width=0.85\columnwidth, clip]{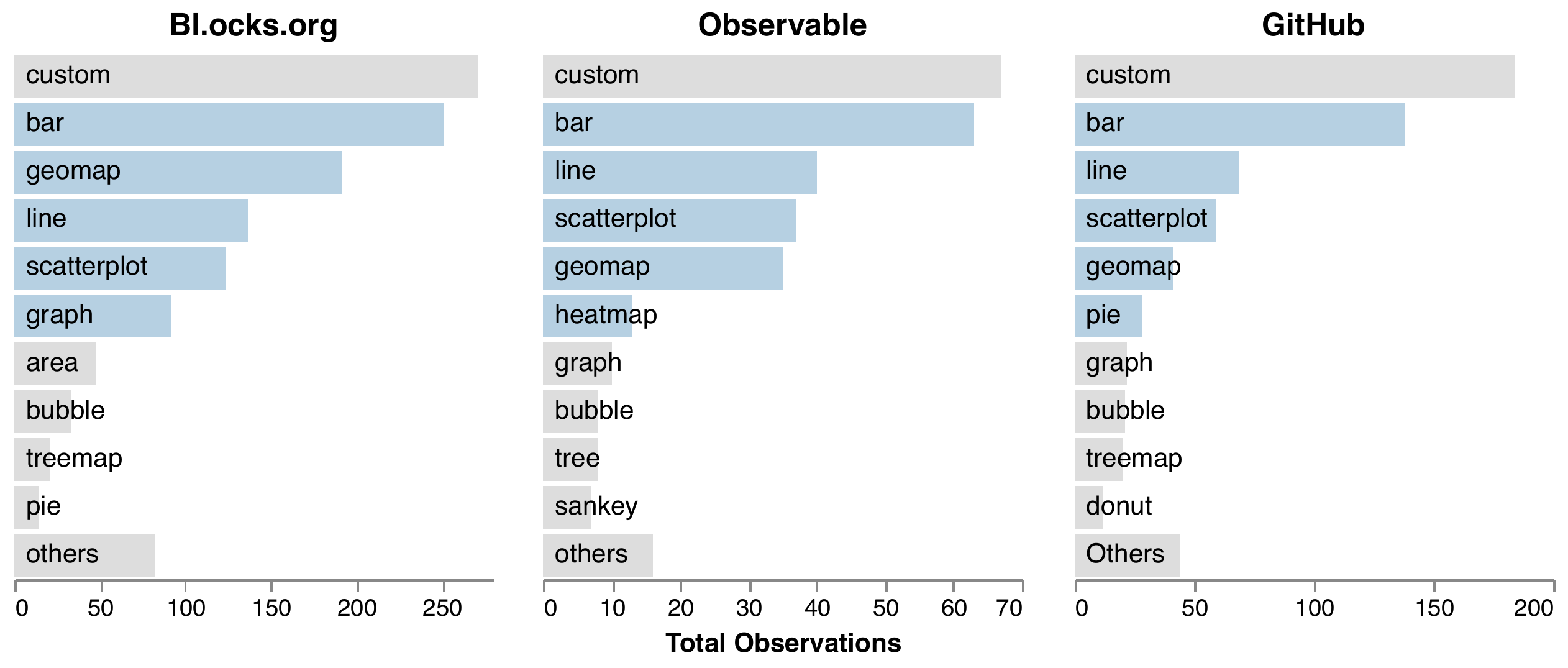}
    \vspace{-3mm}
    \caption{Rankings of observed visualizations in our analysis.}
    \label{fig:viz_types}
\end{subfigure}%
\begin{subfigure}{0.4\linewidth}
    \centering
    \includegraphics[width=0.75\columnwidth, clip]{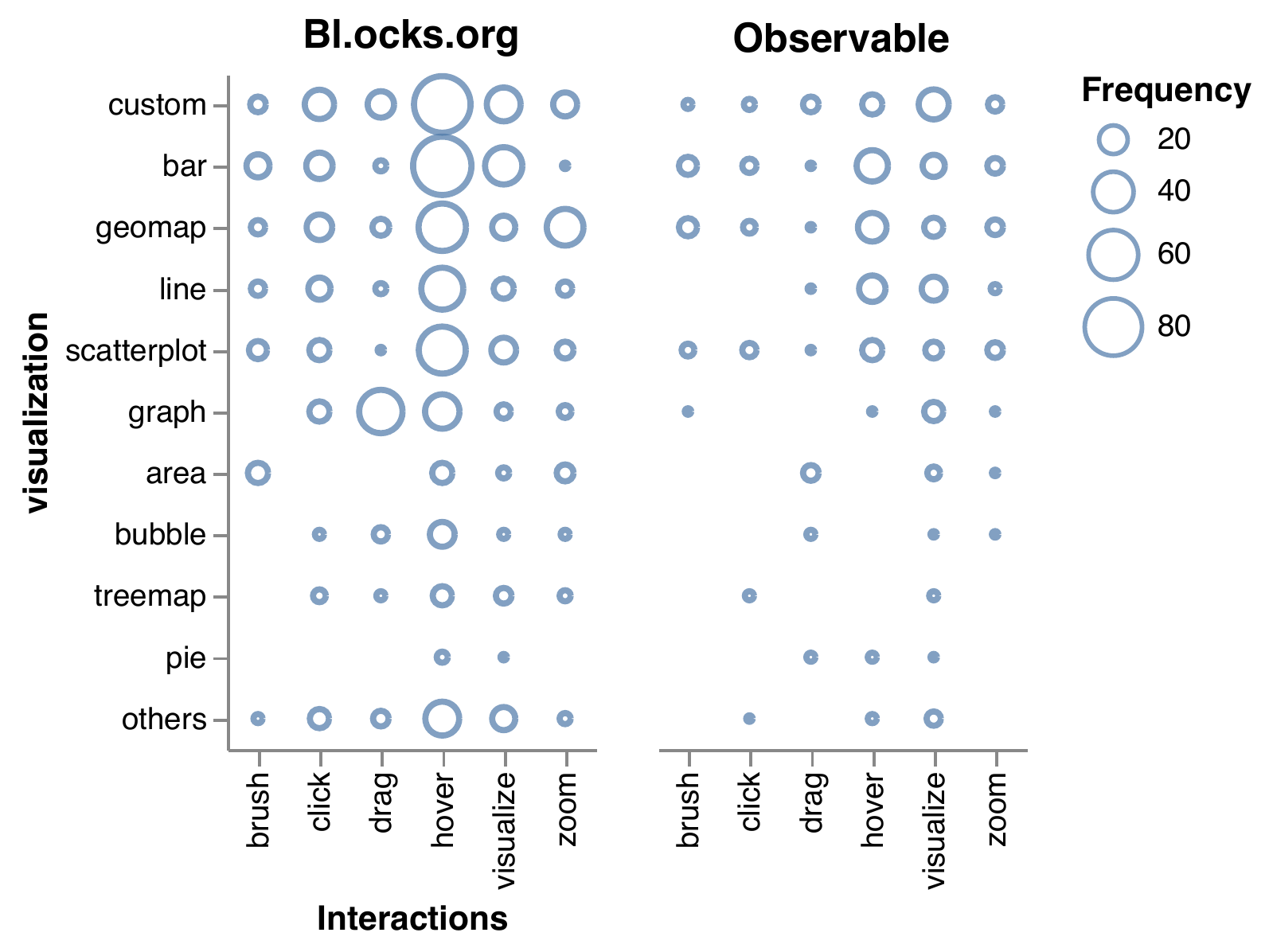}
    \vspace{-3mm}
    \caption{Observed visualization-interaction pairings.}
    \label{fig:int_types}
\end{subfigure}%
\vspace{-4mm}
\caption{Results from analysis carried out on the three example corpus (a) presents the frequency of the top visualization types and their corresponding interactions observed in our datasets. Visualizations with 1\%  and less frequency are captured in the \textit{"others"} category. 
(b) shows the number of times different interaction types were implemented for each visualization type in our corpus}
\label{fig:data_results}
\vspace{-5mm}
\end{figure*}

\vspace{-1mm}
\section{Building a Corpus of D3 Examples}
\vspace{-1mm}

To simplify visualization implementation in D3, we first need to gain a deeper understanding of how users currently use D3.
The first step of our analysis was to collect a corpus of publicly available D3 examples from the internet. We adopted Battle et al.'s approach of identifying ``islands'' or specific websites with thousands of D3 examples~\cite{battle2018beagle}.
We considered three sources, which together represent the most well-known online repositories containing D3 examples: Bl.ocks.org~\cite{blocks:2016}, Observable~\cite{observable:2016} and GitHub~\cite{GitHub:2008}. 
To assess the quality and viability of each dataset for our analysis, we scraped 500 random D3 examples from each corpus and qualitatively analyzed them. To maintain quality within our dataset, a visualization or interaction is only classified if (1) the example explicitly imports D3 \textbf{and} (2) when run, the code produces a visualization in the browser.

We found that the GitHub examples were low quality, due primarily to incomplete code. In addition, unlike the (now deprecated) Bl.ocks.org repository, exported Observable examples are designed to operate within the Observable environment rather than standard web projects, making it more difficult to programmatically analyze them for D3 API usage. Overall, the Bl.ocks.org examples were of consistently higher quality than the GitHub and Observable examples\footnote{The coded examples are available on OSF:
\url{https://osf.io/k58bp/?view_only=72fa3798bbaa4263b5ad662b26a70cb3}.}.
We believe this difference in quality stems from the maturity of the Bl.ocks.org repository compared to Observable, as well as its intended use as a robust repository of D3 examples compared to GitHub.
For these reasons, we analyzed an additional 1000 examples from Bl.ocks.org, resulting in a final corpus with 2500 examples. We report on the visualizations found in all three sources but focus the bulk of our analysis in~\autoref{sec:corpus:design patterns} on the examples from Bl.ocks.org. 

\vspace{-1mm}
\section{What visualization types do D3 users implement?}
\label{sec:corpus:visualizations}
\vspace{-1mm}
We qualitatively coded the corpus by classifying what visualization(s) were implemented in each example according the taxonomy of D3 visualizations observed by Battle et al~\cite{battle2018beagle}. This resulted in 21 visualization types which we refer to as \emph{standard} visualizations. Each distinct visualization found in an example was classified separately. For instance, if both a bar chart and scatterplot were used in a multiple linked view visualization, we classified them separately by visualization type. We discuss the results of our analysis below.

\textbf{Bl.ocks.org Corpus:}
We coded a total of 1500 D3 examples from Bl.ocks.org from which 1265 viable visualizations were found. 994 (78.6\%) of these visualizations were \emph{standard} D3 visualization types such as scatter plots, area charts, voronoi diagrams, etc.. Bar charts (19.8\% n=251), Geographic maps (15.2\% n=192), and line charts (10.8\% n=137) were most prevalent (see \autoref{fig:viz_types}). The top 5 standard visualizations i.e., Bar Charts, Geographic Maps, Line Charts, Scatterplots and Graphs account for 80.1\% of all the standard visualizations implemented with the 16 other visualization types [e.g. Heatmap(1.0\%), Voronoi(0.9\%), Sankey(0.8\%) etc.] accounting for the remaining 19.9\%, as shown in \autoref{fig:viz_types}. We find that  21.4\% (n=271) of these visualizations could not be classified by of the 21 standard visualization types. These examples were diagrams, art, or highly specialized visualizations, such as a departures board for flights, braille clocks, etc., which we classify as \emph{bespoke} visualizations (see~\autoref{fig:bespoke} for examples). 

Our analysis shows that the overwhelming majority of visualizations still conform to the most common visualization types observed on the web. Consistent with prior work~\cite{battle2018beagle}, just 5 visualization types account for the vast majority of these common visualizations. This suggests that we can support the needs of most D3 users by \emph{focusing on the most popular visualizations from our dataset}.

\textbf{Observable Corpus:}
We coded 500 Observable notebooks. However, \(49.2\%\) of these notebooks were unrelated to D3. For example, they did not involve data visualizations or used other visualization tools~\footnote{Please see \url{https://observablehq.com/@thisistaimur/warc-study-analysis} and \url{https://observablehq.com/@aldo/tuftes-charts-in-vega-lite} for examples.}. Of the remaining 254 notebooks, we found 304 viable visualizations, of which 237 (78\%) were standard D3 visualization types comprised of Bar charts (20.7\% n=63), Line charts (13.2\%, n=40), and Scatterplots (12.2\%, n=37). 
The distribution of the Observable visualizations mimicks that of the Bl.ocks.org corpus as the top 5 visualizations account for 79\% of the observed visualizations while 11 visualization types account for the remaining 21\% of the visualizations. However the rankings for the top 5 visualization types are different in both corpora as seen in~\autoref{fig:viz_types}.  

We observed a total of 67 (22\%) \emph{bespoke} visualizations in our corpus. These visualizations range from Hyperboloid plots to Cartograms as seen in~\autoref{fig:bespoke}. We find that the bespoke charts created in Observable were more complicated than those available in other mediums. This may be a result of newer visualization libraries and extensions that are available in the Observable environment.


\textbf{GitHub Corpus:}
500 GitHub repositories were coded in our analysis. While all of these repositories imported the D3 API, only 366 contained valid data visualizations. We found 638 visualizations of which 71.2\% (n=454) were standard D3 visualization types. The most popular visualization types were Bar charts (30.4\% n=138), Line charts (15.2\% n=69) and Scatterplots (13\% n=59). Similar to our other corpora, the top 5 visualizations make up 73.8\% of all standard visualizations on the internet as seen in ~\autoref{fig:viz_types}. We observed a total of 184 (28.8\%) \emph{bespoke} visualizations in our corpus.

\vspace{-1mm}
\section{What interaction types do D3 users implement?}
\label{sec:corpus:interactions}
\vspace{-1mm}

Next, we classify the kinds of interactions users implement in D3, using a similar approach to \autoref{sec:corpus:visualizations}. However, existing interaction taxonomies focus on interface elements rather than code components, requiring us to translate D3 API components into abstract interaction types.
In examining the D3 API and documentation, we extracted 6 common interaction widget types: Brush, Click, Drag, Hover, Visualize, and Zoom. We classify these interactions using Brehmer and Munzer's multi-level typology of visualization tasks~\cite{brehmer2013typology}.
\begin{itemize}[nosep]
    \item \textbf{Select:} highlighting data points to emphasize salient information, often using ``\textit{Brush}'', ``\textit{Hover}'', and ``\textit{Click}'' widgets.
    \item \textbf{Encode:} changing which data attributes are encoded in a visualization, often using a GUI widget; we use the term \textit{``Visualize''} to represent all such encoding interactions.
    \item \textbf{Navigate:} Re-centering a visualisation on a specific subset of points and granularity, often using ``\textit{Zoom \& Pan}'' widget.
    \item \textbf{Arrange:} sorting and organizing marks within the visualization, such as by using a ``\textit{Drag}'' widget.
\end{itemize}
In general we observe a high number of interactions being implemented within visualizations suggesting that users may follow recommended best practices to enable interaction when programming their visualizations~\cite{yi2007roleofinteractions}. We discuss the details of our analysis for the Bl.ocks.org and Observable corpus below~\footnote{Interactions were not coded for GitHub due to data quality issues.}.

\textbf{Bl.ocks.org Corpus}
Of the 1265 viable examples coded, 659 (52.1\%) contained interactions. 859 interaction implementations were identified within these examples, with the most popular widget types being Hover (n=390), Visualize (n=118) and Click (n=100). We observed 62.6\% of all interactions were used to ``\textit{Brush}'', ``\textit{Hover}'', and ``\textit{Click}'' specific data points in visualizations. Mapping these observations to the Brehmer and Munzer typology, we find \textbf{selection} to be the main form of interaction used, being implemented  in 62.6\% of D3 visualizations in our analysis. 13.7\% of implemented interactions \textbf{encode} new data attributes, 13.2 \% support \textbf{navigation} through visualizations and 10.5\% to \textbf{arrange} marks on the screen using the \textit{``Drag''} widget. 
Our findings also revealed that certain interaction types were often implemented \emph{together} in the examples. Examining these associations, we identify 39 distinct pairs of interactions with \textit{``Click and Hover''} being the most frequent pairing representing 14\% of all occurrences.

\textbf{Observable Corpus}
We had a total of 304 visualization examples in our Observable corpus, of which 48.2\% were interactive. 42\% of all interactions represented \textbf{selections} and a total of 174 interactions were identified with \textbf{encoding} new data attributes using the \textit{Visualize} widget being used 35.6\%  (n=62) of the time. 12.1\% with \textbf{navigating} through visualizations, and 10.3\% with \textbf{arranging} visual elements in visualizations. 58 of these interactions (33.3\%) were implemented in pairs with 12.1\% belonging to the \textit{``Visualize and Hover''} pairs. 
\begin{figure*}[h!]
 \centering 
 \begin{subfigure}{0.33\linewidth}
    \centering
    \includegraphics[width=0.9\columnwidth, trim=30mm 2mm 30mm 2mm, clip, scale=1.5]{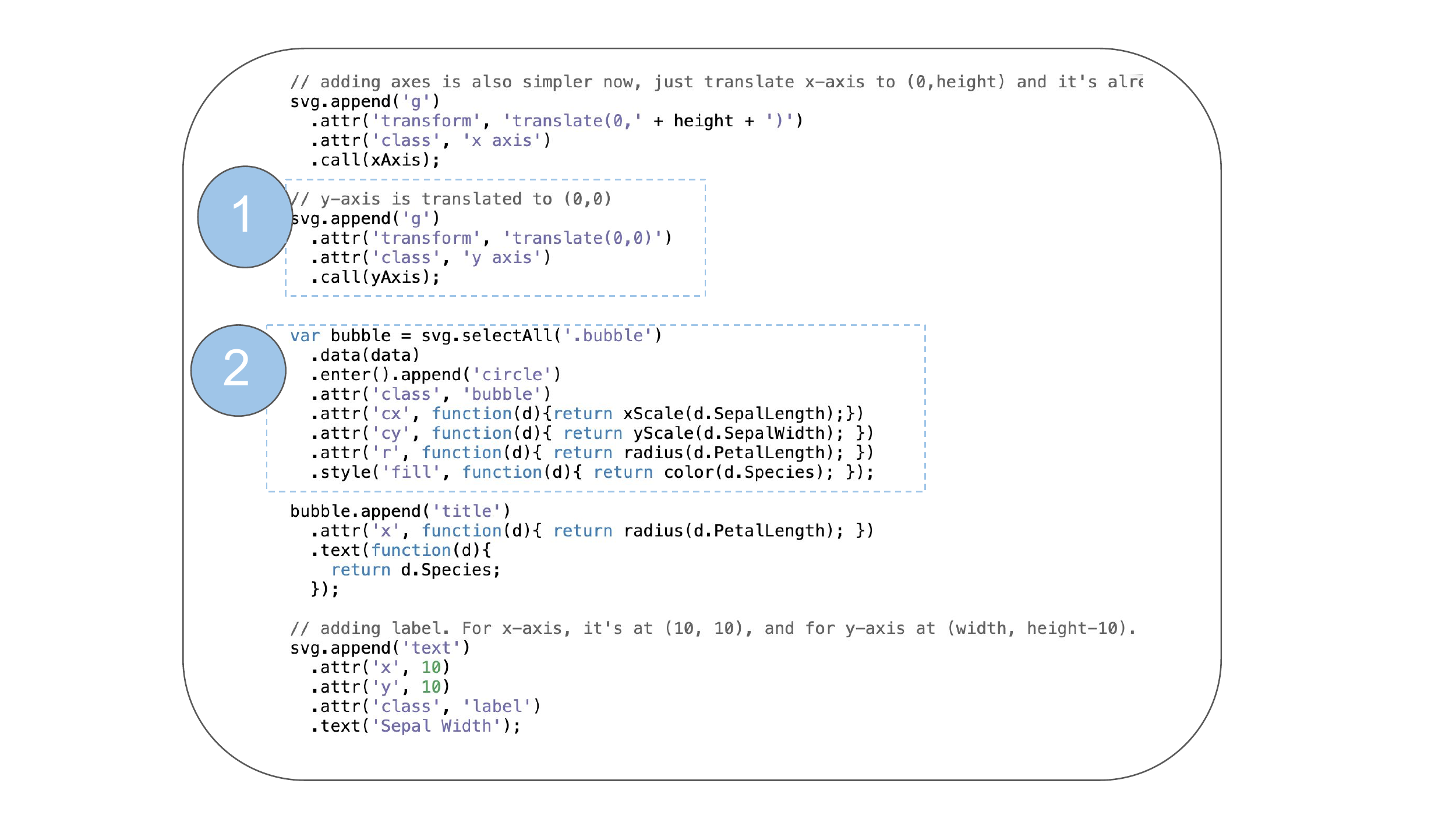}
    \vspace{-3mm}
    \caption{Scatterplot Example}
    \label{fig:scatter_example1}
\end{subfigure}%
\begin{subfigure}{0.33\linewidth}
    \centering
    \includegraphics[width=0.9\columnwidth, trim=25mm 2mm 25mm 2mm, clip]{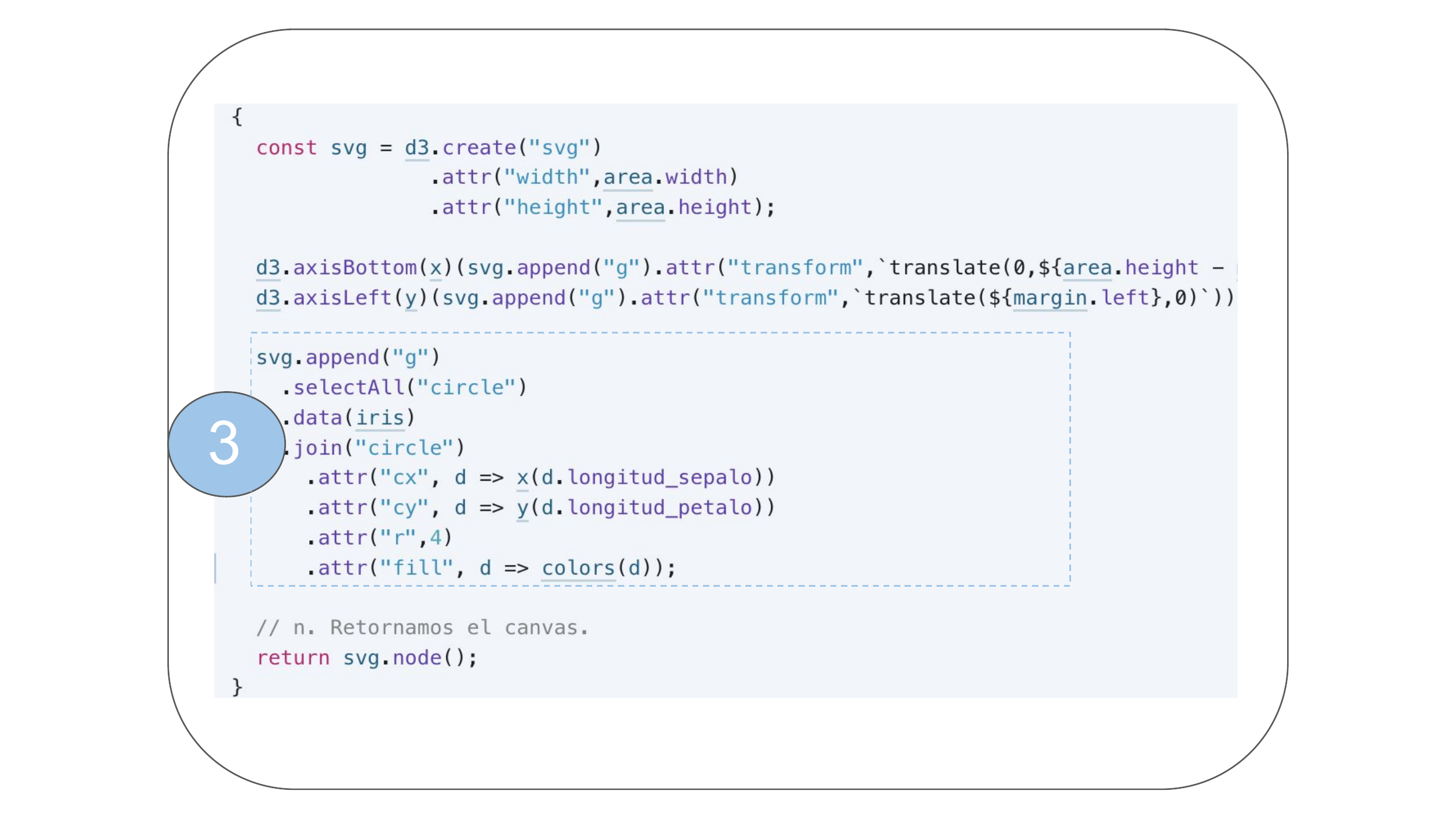}
    \vspace{-3mm}
    \caption{Scatterplot Example}
    \label{fig:scatter_example2}
\end{subfigure}%
\begin{subfigure}{0.33\linewidth}
    \centering
    \includegraphics[width=0.9\columnwidth, trim=20mm 2mm 24.5mm 2mm, clip]{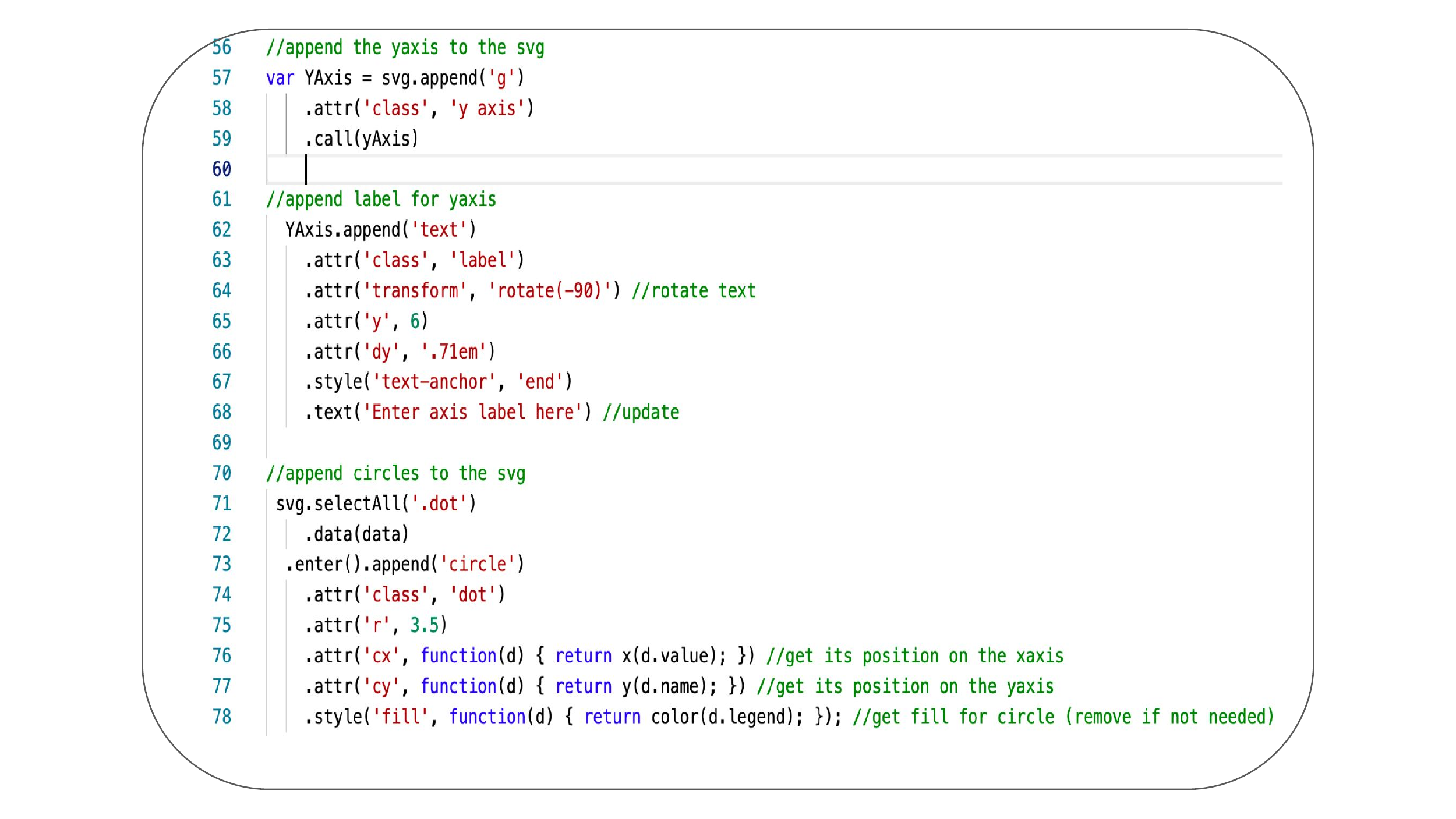}
    \vspace{-3mm}
    \caption{Generic Template}
    \label{fig:generic_template}
\end{subfigure}
 \vspace{-3mm}
 \caption{Reoccurring code blocks (1, 2 \& 3) in two scatterplot examples (A \& B) are synthesized into a generic scatterplots template (C).}
 \label{fig:templatesynthesis}
 \vspace{-5mm}
\end{figure*}

\vspace{-1mm}
\section{How are these visualizations being Implemented?}
\label{sec:corpus:design patterns}
\vspace{-1mm}

A key step in clarifying the needs of D3 users is to understand \emph{how} they implement visualizations.
For each visualization and interaction type, we examined multiple user implementations to understand: the syntactical correctness of each example, the organization and structure of the code, and the API calls used. 

\vspace{-1mm}
\subsection{Code Structure} 
\label{sec:corpus:design patterns:code structure}
\vspace{-1mm}
At the low level, users' programs vary slightly for visualizations of the same type, such as by variable names, white space used, and so on.  However, when the code is examined at the structural level, the overall order and API calls remain the same. For example, the default code order across visualizations begins with defining the dimensions for the visualization (i.e. width, height, and margins) and creating an SVG object where subsequent visualization elements will be housed. 
The baseline code for specific visualization types follow almost the same structure. For example, the highlighted code blocks in~\autoref{fig:scatter_example1} and~\autoref{fig:scatter_example2}  highlight similar code structures for defining point marks in two different scatterplot implementations, with slight differences to fit users' datasets. Several examples even contained direct links to other repositories as inspiration. Oftentimes, multiple lines of code were copied directly from these reference repositories. This corroborates past observations on code copying as a common implementation strategy in the D3 community~\cite{hogue2020searchingd3, battle2021exploring}. 

\vspace{-1mm}
\subsection{Use of Interactions across Visualization Types}
\label{sec:corpus:design patterns:interactions}
\vspace{-1mm}
To expand upon our observation, we measure how often popular interaction widgets are implemented for each visualization type.


\textbf{Some interaction widgets appear to be universal.} We found that certain widget types were pervasive across most visualizations. For example, hovering is the most common interaction widget for nearly every visualization type observed in \autoref{fig:int_types}. These findings suggest that certain widget types may be considered universal, hinting at code blocks that may be strong candidates for reuse.

\textbf{The number of interaction widgets varies across visualization types.}  For certain visualization types (e.g. Graphs) including at least one widget was the convention. For example, 49\% of the line chart examples implemented had no interaction widgets, whereas 82\% of the graph examples had at least one widget implemented.

\textbf{Users implement interaction widgets according to visualization type.} We find that the types of interaction widgets implemented varies by visualization type (see~\autoref{fig:int_types}).
For example, we see in \autoref{fig:data_results} that line charts have a wider set of implemented interaction widgets compared to area charts, where hovering was more prevalent in line charts and brushing in area charts. We also see that geographic maps are more likely to have zooming implemented compared to other visualization types, and graphs are more likely to support dragging.
Furthermore, the set of interactions implemented were dependent on the visualization type.
For example, brushing and hovering were often implemented together in scatterplots.

These findings indicate that for certain D3 implementations, users could benefit from multiple modes of interaction to explore a visualization.
For example, the user may (for graphs) or may not (for bar charts) need to consider incorporating interactions and may want to use specific sets of interactions to match existing D3 examples.

\vspace{-1mm}
\subsection{Generating Reusable D3 Templates}
\label{sec:generating_code}
\vspace{-1mm}
Given an understanding of the most common visualizations (\autoref{sec:corpus:visualizations}) and interactions (\autoref{sec:corpus:interactions}) implemented in D3, we sought to ease the burden of users in implementing these visualizations. To do this, we randomly selected examples of eight common visualization types and translated common implementation patterns (\autoref{sec:corpus:design patterns:code structure}) into general-purpose code templates. While publicly available D3 templates exist on platforms like D3Live~\cite{d3.js/live}, they are curated based on the template designer's own understanding of D3. Our user-driven template synthesis approach encapsulates common programming practices from \emph{hundreds} of D3 users into a single template.
 
D3 API calls used within examples remain consistent for visualization types. For example, scatterplots and bubble charts consistently contain calls to create circle marks. We rely on D3's API structure and our observations of common D3 code structures to infer the purpose of each line of code and the role they play in creating a visualization.
Through this analysis, we manually extracted common code sections for each visualization type.
Using scatterplots as an example, we identified common code sections for creating circle marks as shown in~\autoref{fig:scatter_example1} \&~\autoref{fig:scatter_example2}. These code sections were combined to generate a working generic implementation of each visualization that can be modified to fit a user's dataset (see~\autoref{fig:generic_template}). These generic implementations can be used as templates~\footnote{ \label{osf} The curated templates are available on OSF: \url{https://osf.io/k58bp/?view_only=72fa3798bbaa4263b5ad662b26a70cb3}}.

\vspace{-1mm}
\section{Discussion}
\vspace{-1mm}
Using our insights into how D3 users implement visualizations, previous work, and our own experiences working with D3, we provide design considerations for simplifying the visualization programming process with visualization languages.



\noindent\textbf{C1: Maximize code component reuse:} Code reuse is a documented and advocated skill within the programming and D3 community~\cite{hogue2020searchingd3, battle2021exploring}. Our analysis in \autoref{sec:corpus:design patterns} reveals striking structural similarities for various visualization types. Users tend to implement D3 visualizations in consistent ways, where these practices are largely drawn from existing examples.
However, recent literature also finds that users often struggle to reuse existing D3 examples because they are typically \emph{not modularized}, making it hard to extract only the relevant functionality~\cite{battle2021exploring}.
We could speed up the implementation process by automating the adoption of established coding conventions across visualization and interaction types. Based on this premise, we synthesize code \emph{templates} from existing D3 examples to capture existing implementation patterns among D3 users, which could be adopted in future visualization tools.
Furthermore, we annotate major code blocks within each template with their intended purpose. This makes it easy for users to quickly associate a code segment in the flow of a template with the specific task it performs.



\noindent\textbf{C2: Use usage statistics to drive adaptive recommendations:} Our analysis in \autoref{sec:corpus:design patterns} shows that a high fraction of D3 examples allow their end users to explore the underlying data through interactions, reinforcing the importance of interaction in visualizations~\cite{dimara2020interactions}.
In contrast, we see an emphasis on recommending visualization designs but not interaction designs in most visualization recommendation and automated design systems~\cite{zeng2021evaluation}.
We observed many combinations of visualizations and interactions. As a result, D3 users interested in implementing interactions would have to sift through numerous examples to identify the appropriate implementation conventions for their target visualization type. To save time and effort, we can instead automatically recommend a curated list of appropriate interactions to implement for a given visualization type. 

\noindent\textbf{C3: Involve the community in tool refinement:} The large and active D3 community provides a unique opportunity to \emph{co-design} effective visualization tools. For example, recommendation models could be improved by combining visualization heuristics and reinforcement learning to actively learn from the behaviour of D3 users. In terms of interactions, recommenders could infer the current visualization context from the user's code and and compare it with similar visualizations created in the past to provide customized recommendations for potential interactions to implement. Community members could even take a more active role in the tuning of recommenders by providing direct guidance, e.g., annotations, for what the underlying models \emph{should learn} (e.g., code modularization, perceptually effective encoding parameters) or \emph{should not learn} (e.g., poor coding conventions or encoding decisions) from existing examples.

\vspace{-1mm}
\subsection{Limitations}
\vspace{-1mm}
Our work relies on the efforts of thousands of D3 users who publish their work on D3 forums online. Our results are representative of the this population of users but may not generalize to other contexts of D3 usage such as those who use D3 to implement visualizations for news articles or internal usage within organizations. Furthermore, our analysis of the interactions in D3 focuses mainly on interactions that are supported by the native D3 API. We acknowledge that this potentially excludes other interactions that are supported by external modules and libraries, thereby limiting the richness of interaction types found in our study.
However, our analysis techniques can easily be applied to other languages, modules and libraries for interactive visualization programming (e.g., Vega-Lite~\cite{satyanarayan2017vega-lite}).
We also encourage the community to conduct studies in the future to verify how and whether templates simplify the visualization design process.

\vspace{-1mm}
\section{Conclusion and Future Work}
\label{sec:conclusion}
\vspace{-1mm}

In this paper, we investigate visual and interactive design patterns implemented by D3 users through 2500 examples shared in three online repositories (Bl.ocks.org, Observable, GitHub). Five visualization types accounted for over 80\% of all the D3 visualizations we observed.
We observed that users generally used the same strategies to implement the same visualization type.
Furthermore, certain interaction types were frequently paired with specific visualization types.
We extracted the common implementation strategies we observed and synthesized templates representing universal building blocks for implementing the most popular D3 visualizations and interactions. These templates provide a baseline that D3 users can build upon to program new visualizations with less time and effort.
\vspace{-0.5mm}


\acknowledgments{
\vspace{-0.5mm}
This work was supported in part by NSF award IIS-1850115 and a VMWare Early Career Faculty Grant.
}

\bibliographystyle{abbrv-doi}

\bibliography{references}
\end{document}